\documentclass[sigconf,anonymous = false]{acmart}
\usepackage[export]{adjustbox}
\usepackage[normalem]{ulem}
\usepackage{xcolor}

\usepackage{dirtytalk}
\usepackage{color-edits}
\addauthor{bh}{violet}
\addauthor{mm}{blue}

\AtBeginDocument{%
  \providecommand\BibTeX{{%
    \normalfont B\kern-0.5em{\scshape i\kern-0.25em b}\kern-0.8em\TeX}}}

\setcopyright{acmcopyright}
\copyrightyear{2020}
\acmYear{2020}
\setcopyright{rightsretained}
\acmConference[FAT* '20]{Conference on Fairness, Accountability, and Transparency}{January 27--30, 2020}{Barcelona, Spain}
\acmBooktitle{Conference on Fairness, Accountability, and Transparency (FAT* '20), January 27--30, 2020, Barcelona, Spain}\acmDOI{10.1145/3351095.3372873}
\acmISBN{978-1-4503-6936-7/20/02}

\begin{document}

%%
%% The "title" command has an optional parameter,
%% allowing the author to define a "short title" to be used in page headers.
\title{Closing the AI Accountability Gap:\\ Defining an End-to-End Framework for Internal Algorithmic Auditing}

%%
%% The "author" command and its associated commands are used to define
%% the authors and their affiliations.
%% Of note is the shared affiliation of the first two authors, and the
%% "authornote" and "authornotemark" commands
%% used to denote shared contribution to the research.
\author{Inioluwa Deborah Raji}
\affiliation{%
 \institution{Partnership on AI}}
\email{deb@partnershiponai.org}
\authornote{Both authors contributed equally to this paper. This work was done by Inioluwa Deborah Raji as a fellow at Partnership on AI (PAI), of which Google, Inc. is a partner. This should not be interpreted as reflecting the official position of PAI as a whole, or any of its partner organizations.}
\author{Andrew Smart}
\authornotemark[1]
\affiliation{%
 \institution{Google}}
\email{andrewsmart@google.com}

\author{Rebecca N. White}
\affiliation{%
  \institution{Google}}

\author{Margaret Mitchell}
\affiliation{%
  \institution{Google}}
  
\author{Timnit Gebru}
\affiliation{%
  \institution{Google}}

\author{Ben Hutchinson}
\affiliation{%
  \institution{Google}}
  
\author{Jamila Smith-Loud}
\affiliation{%
  \institution{Google}}

\author{Daniel Theron}
\affiliation{%
  \institution{Google}}
  
\author{Parker Barnes}
\affiliation{%
  \institution{Google}}

%%
%% This command processes the author and affiliation and title
%% information and builds the first part of the formatted document.
%\maketitle

%%
%% By default, the full list of authors will be used in the page
%% headers. Often, this list is too long, and will overlap
%% other information printed in the page headers. This command allows
%% the author to define a more concise list
%% of authors' names for this purpose.
\renewcommand{\shortauthors}{Raji \& Smart, et al.}

%%
%% The abstract is a short summary of the work to be presented in the
%% article.
\begin{abstract}

%The use of machine learning systems in consequential domains is rapidly increasing.

Rising concern for the societal implications of artificial intelligence systems has inspired a wave of academic and journalistic literature in which deployed systems are audited for harm by investigators from outside the organizations deploying the algorithms. However, it remains challenging for practitioners to identify the harmful repercussions of their own systems prior to deployment, and, once deployed, emergent issues can become difficult or impossible to trace back to their source. 

%Audits have been shown to be more effective in leading to accountability when adhering to established and transparent protocols. As such, i

In this paper, we introduce a framework for algorithmic auditing that supports artificial intelligence system development end-to-end, to be applied throughout the internal organization development lifecycle. Each stage of the audit yields a set of documents that together form an overall audit report, drawing on an organization's values or principles to assess the fit of decisions made throughout the process. 
The proposed auditing framework is intended to contribute to closing the {\it accountability gap} in the development and deployment of large-scale artificial intelligence systems by embedding a robust process to ensure audit integrity.

%We incorporate approaches from social science, responsible innovation practice, and other safety-critical industries to create an audit process aimed at anticipating harms, hazards, and potential biases before a system is released. The audit framework comprises multiple optional stages, enabling customization depending on the scale, impact, and context of the project. 

%\mmdelete{ We then discuss the inherent limitations of internal audits, and explain how they are just one component of a suite of accountability measures required to ensure responsible machine learning deployment.}
% 
% 
\end{abstract}

%%
%% The code below is generated by the tool at http://dl.acm.org/ccs.cfm.
%% Please copy and paste the code instead of the example below.
%%
\begin{CCSXML}
<ccs2012>
<concept>
<concept_id>10003456.10003457.10003490.10003507</concept_id>
<concept_desc>Social and professional topics~System management</concept_desc>
<concept_significance>500</concept_significance>
</concept>
<concept>
<concept_id>10003456.10003457.10003490.10003507.10003509</concept_id>
<concept_desc>Social and professional topics~Technology audits</concept_desc>
<concept_significance>500</concept_significance>
</concept>
<concept>
<concept_id>10011007.10011074.10011081</concept_id>
<concept_desc>Software and its engineering~Software development process management</concept_desc>
<concept_significance>500</concept_significance>
</concept>
</ccs2012>
\end{CCSXML}

\ccsdesc[500]{Social and professional topics~System management}
\ccsdesc[500]{Social and professional topics~Technology audits}
\ccsdesc[500]{Software and its engineering~Software development process management}

%%
%% Keywords. The author(s) should pick words that accurately describe
%% the work being presented. Separate the keywords with commas.
\keywords{Algorithmic audits, machine learning, accountability, responsible innovation}

%%
%% This command processes the author and affiliation and title
%% information and builds the first part of the formatted document.
\maketitle

%% A "teaser" image appears between the author and affiliation
%% information and the body of the document, and typically spans the
%% page.
%\begin{teaserfigure}
%  \includegraphics[width=\textwidth]{sampleteaser}
%  \caption{Seattle Mariners at Spring Training, 2010.}
%  \Description{Enjoying the baseball game from the third-base
%  seats. Ichiro Suzuki preparing to bat.}
%  \label{fig:teaser}
%\end{teaserfigure}

\section{Introduction}
With the increased access to artificial intelligence (AI) development tools and Internet-sourced datasets, corporations, nonprofits and governments are deploying AI systems at an unprecedented pace, often in massive-scale production systems impacting millions if not billions of users \cite{al2015efficient}. In the midst of this widespread deployment, however, come valid concerns about the effectiveness of these automated systems for the full scope of users, and especially a critique of systems that have the propensity to replicate, reinforce or amplify %certain 
harmful existing social biases \cite{buolamwini2018gender, raji2019actionable, kiritchenko2018examining}. External audits are designed to identify these risks from outside the system and serve as accountability measures for these deployed models. However, such audits tend to be conducted after model deployment, when the system has already negatively impacted users \cite{green2019disparate, moy2019police}. %To minimize the harm done to users, organizations building and deploying AI systems that have the potential to impact society must work together with external stakeholders to ensure that risks to society and rights are anticipated and taken into consideration during the development and maintenance of large-scale AI systems. %Once deployed, these systems must be continuously measured for their alignment with  ethical priorities declared prior to deployment. 

%\mmcomment{Is it important that they be declared prior to deployment? What is new ethical priorities emerge post-deployment?  Does it make sense to remove that last bit?}

\begin{figure}[b]
  \centering
    \includegraphics[width=0.5\textwidth, right]{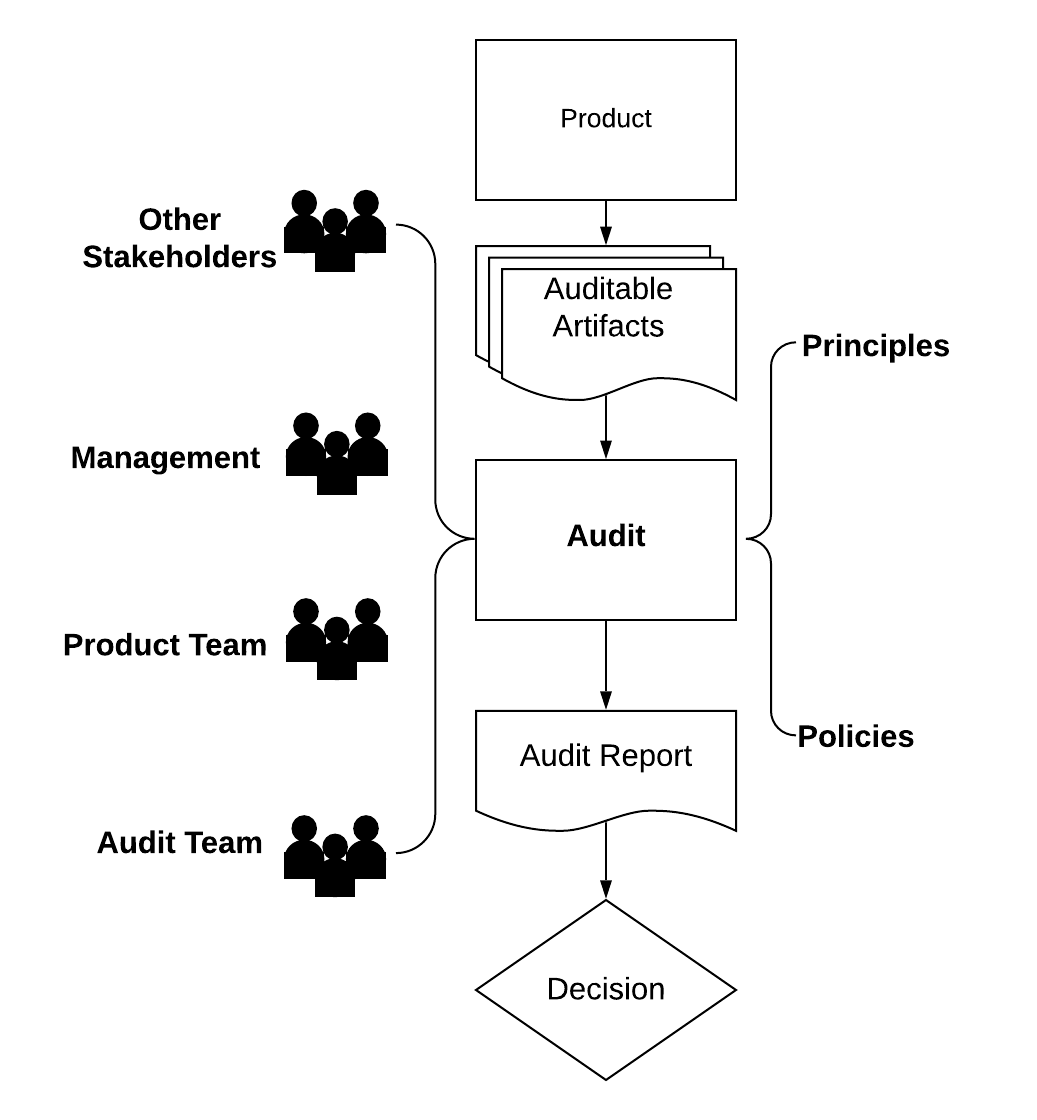}
  \caption{High-level overview of the context of an internal algorithmic audit. The audit is conducted during product development and prior to launch. The audit team leads the product team, management and other stakeholders in contributing to the audit. Policies and principles, including internal and external ethical expectations, also feed into the audit to set the standard for performance. 
 }
 \label{fig:audit1}
\end{figure}

%---sketched in Figure~\ref{fig:audit1}---

%This paper provides an internal algorithmic auditing framework to assist organizations in adhering to declared AI principles.
%Our framework provides much needed structure to ML engineering for ethical objectives. 

%Those involved in the algorithmic development process need to look to more interdisciplinary strategies - from social science, anthropology, finance, medicine and business - in addition to other engineering domains, in order to appropriately capture the diverse nature of concerns. 

In this paper, we present internal algorithmic audits as a mechanism to check that the engineering processes involved in AI system creation and deployment meet declared ethical expectations and standards, such as organizational AI principles. The audit process is necessarily boring, slow, meticulous and methodical---antithetical to the typical rapid development pace for AI technology. However, it is critical to slow down as algorithms continue to be deployed in increasingly high-stakes domains. By considering historical examples across industries, we make the case that such audits can be leveraged to anticipate potential negative consequences before they occur, in addition to providing decision support to design mitigations, more clearly defining and monitoring potentially adverse outcomes, and anticipating harmful feedback loops and system-level risks \cite{ensign2017runaway}. Executed by a dedicated team of organization employees, internal audits operate within the product development context and can inform
%decisions to say "no", i.e., 
% potentially informing 
the ultimate decision
to abandon the development of AI technology when the risks outweigh the benefits (see  Figure~\ref{fig:audit1}).

%As an established audit process is key to maintaining audit integrity \cite{tax_legit}, 
Inspired from the practices and artifacts of several disciplines, we go further to develop SMACTR, a defined internal audit framework meant to guide practical implementations. Our framework strives to establish interdisciplinarity as a default in audit and engineering processes while providing the much needed structure to support the conscious development of AI systems.

%After making the case for internal audits, we then outline a proposed framework as one mechanism for ethical compliance through auditing prior to deployment. %In the absence of industry consensus on universal standards, the framework makes use of customized ethical principles \cite{zeng2018linking,jobin2019artificial, mittelstadt2016ethics} as a proxy of the prioritized risks and ethical expectations involved in ML model deployment within a specific industry context. %Given an example of an AI principle statement, we walk through a hypothetical example of the proposed audit procedure and present the corresponding requirements for execution as an aid in ethical decision making around key choices towards more responsible engineering development prior to deployment.

\section{Governance, Accountability and Audits}

We use \textit{accountability} to mean the state of being responsible or answerable for a system, its behavior and its potential impacts \cite{kohli2018translation}.
Although algorithms themselves cannot be held accountable as they are not moral or legal agents \cite{bryson2017and}, the organizations designing and deploying algorithms can through \textit{governance} structures. Proposed standard ISO 37000 defines this structure as "the system by which the whole organization is directed, controlled and held accountable to achieve its core purpose over the long term."\footnote{https://committee.iso.org/sites/tc309/home/projects/ongoing/ongoing-1.html} If the responsible development of artificial intelligence is a core purpose of organizations creating AI, then a governance system by which the whole organization is held accountable should be established.
 
In environmental studies, Lynch and Veland \cite{lynch2018urgency} introduced the concept of \textit{urgent governance}, % to describe the problem of governance for  the climate crisis. %We suggest that algorithmic governance also requires urgency, as artificial intelligence systems are both scientifically and technically complex, and also becoming increasingly embedded in the structure of social and economic institutions, with human-consequential implications. 
%In the environmental context, there is the known distinction 
distinguishing between \textit{auditing} for system reliability vs societal harm. For example, 
a power plant can be consistently productive while causing harm to the environment through pollution \cite{leveson2011engineering}. Similarly, an AI system can be found technically reliable and functional through a traditional engineering quality assurance pipeline without meeting declared ethical expectations. A separate governance structure is necessary for the evaluation of these systems for ethical compliance. This evaluation can be embedded in the established quality assurance workflow but serves a different purpose, evaluating and optimizing for a different goal centered on social benefits and values rather than typical performance metrics such as accuracy or profit \cite{kroll2016accountable}. Although concerns about reliability are related, and although practices for testing production AI systems are established for industry practitioners \cite{breck2017ml}, issues involving social impact, downstream effects in critical domains, and ethics and fairness concerns are not typically covered by concepts such as technical debt and reliability engineering.

%This paper provides an internal algorithmic auditing framework to assist organizations in adhering to declared AI principles.
%Our framework provides much needed structure to engineering towards ethical objectives.The audit process is necessarily boring, slow, meticulous and methodical---antithetical to the typical rapid development pace for AI technology. However, it is critical to slow down as algorithms continue to be deployed in increasingly high-stakes domains. 
%Those involved in the algorithmic development process need to look to more interdisciplinary strategies - from social science, anthropology, finance, medicine and business - in addition to other engineering domains, in order to appropriately capture the diverse nature of concerns. 
%Our proposed audit framework takes inspiration from social science, anthropology, finance, medicine and business in addition to other engineering domains, striving to establish interdisciplinarity as a default in internal audit practices and engineering processes.

\subsection{What is an audit?}
\textit{Audits} are tools for interrogating complex processes, often to determine whether they comply with company policy, industry standards or regulations \cite{liu2012enterprise}. The IEEE standard for software development defines an audit as \say{an independent evaluation of conformance of software products and processes to applicable regulations, standards, guidelines, plans, specifications, and procedures} \cite{4601584}. 
Building from methods of external auditing in investigative journalism and research \cite{diakopoulos2014algorithmic, sandvig2014auditing, raji2019actionable}, algorithmic auditing has started to become similar in spirit to the well-established practice of bug bounties, where external hackers are paid for finding vulnerabilities and bugs in released software \cite{maillart2017given}. These audits, modeled after intervention strategies in information security and finance~\cite{raji2019actionable}, have significantly increased public awareness of algorithmic accountability. %and have resulted in some companies making changes 
%and improvements to their APIs due to audit design decisions .% \mmcomment{This says that algorithmic auditing originated from investigative journalism/research AND that they're modelled after intervention strategies in information security and finance.  Tie together.}

%One transformative external audit of commercially available facial analysis systems, 
An external audit of automated facial analysis systems exposed high disparities in error rates among darker-skinned women and lighter-skinned men~\cite{buolamwini2018gender}, showing how structural racism and sexism can be encoded and reinforced through AI systems.
~\cite{buolamwini2018gender} reveals \textit{interaction failures},
in which the production and deployment of an AI system interacts with unjust social structures to contribute to biased predictions, as Safiya Noble has described \cite{noble2018algorithms}.  
Such findings demonstrate the need for companies to understand the social and power dynamics of their deployed systems' environments, and record such insights to manage their products' impact. 

\subsection{AI Principles as Customized Ethical Standards}
%To date, there are multiple industry standards under development for AI governance~\cite{morley2019overview}. 
%However, these incomplete standards are not yet widely adopted by the tech industry or practitioners. 
%These statements of principles can form the basis of a sanity check of ethical compliance prior to model deployment.  
%If a statement of ethical principles can be a critical first step toward responsible AI, the next important step is operationalizing ethical principles. 
According to Mittelstadt \cite{mittelstadt_ai_2019}, at least 63 public-private initiatives have produced statements describing high-level principles, values and other tenets to guide the ethical development, deployment and governance of AI. 
Important values such as ensuring AI technologies are subject to human direction and control, and avoiding the creation or reinforcement of unfair bias, have been included in many organizations' ethical charters. 
However, the AI industry lacks proven methods to translate principles into practice~\cite{mittelstadt_ai_2019}, and AI principles have been criticized for being vague and providing little to no means of accountability~\cite{whittlestone2019role, greene2019better}.  
Nevertheless, such principles are becoming common methods to define the ethical priorities of an organization and thus the operational goals for which to aim~\cite{zeng2018linking, jobin2019artificial}. 
Thus, in the absence of more formalized and universal standards, they can be used as a North Star to guide the evaluation of the development lifecycle, and internal audits can investigate alignment with declared AI principles prior to model deployment.
%Strong internal controls and accompanying audits are essential to maintaining human oversight over these systems, as well as preventing AI systems from failing to reach these ethical expectations. 
We propose a framing of risk analyses centered on the failure to achieve AI principle objectives, outlining an audit practice that can begin translating ethical principles into practice. 
%One purpose of an audit is to provide learnings for a governance framework that could standardize best practices and establish a broader perspective on ethics and ML fairness inside a public or private organization developing AI.

\subsection{Audit Integrity and Procedural Justice}

Audit results are at times approached with skepticism since they are reliant on and vulnerable to human judgment. %For audits to serve as reliable evaluations, they must be externally legitimized through a standardized process \cite{tax_legit}. 
%An emphasis on solidifying organization-wide processes backed by the necessary stakeholders can help ensure independent results that are implemented.
To establish the integrity of the audit itself as an independently valid result, the audit must adhere to the proper execution of an established audit process. This is a repeatedly observed phenomenon in tax compliance auditing, where several international surveys of tax compliance demonstrate that a fixed and vetted tax audit methodology is one of the most effective strategies to convince companies to respect audit results and pay their full taxes~\cite{tax_malay, tax_aust}. 

Procedural justice implies the legitimacy of an outcome due to the admission of a fair and thorough process. Establishing procedural justice to increase compliance is thus a motivating factor for establishing common and robust frameworks through which independent audits can demonstrate adherence to standards. 
In addition, audit integrity is best established when auditors themselves live up to an ethical standard, vetted by adherence to an expected code of conduct or norm in how the audit is to be conducted. In finance, for example, it became clear that any sense of dishonesty or non-transparency in audit methodology would lead audit targets to dismiss rather than act on results \cite{fin_ethics}.

\subsection{The Internal Audit}

External auditing, in which companies are accountable to a third party \cite{raji2019actionable}, are fundamentally limited by lack of access to internal processes at the audited organizations. Although external audits conducted by credible experts are less affected by organization-internal considerations, external auditors can only access model outputs, for example by using an API \cite{sandvig2014auditing}. Auditors do not have access to intermediate models or training data, which are often protected as trade secrets \cite{burrell2016machine}.
%which, for example, can prevent hostile actors from manipulating online services.
%Recent works that have outlined methods for 
%, however, there is little to no guidance on applying and evaluating ethical principles relevant to a product release that can affect millions of people. Our hope is that the framework introduced here can be a complement to external audits, providing additional context and information on the end-to-end system development process. 
%  they are also limited in their access and often unable to impact product development prior to release.  
Internal auditors' direct access to systems can thus help extend traditional external auditing paradigms by incorporating additional information typically unavailable for external evaluations to reveal previously unidentifiable risks. %on how the system works, exposing to the potential issues feeding into the system, increasing the level of accountability for system creators.
%Previous work has identified several strategies for more open and transparent algorithmic design processes to combat bias. For example,  \cite{talend} proposes a system to uncover algorithmic bias across the categories of {\it data bias}, {\it human bias} and {\it algorithmic processing bias}, utilizing a proactive auditing process applied throughout AI development.

The goals of an internal audit are similar to quality assurance, with the objective to enrich, update or validate the risk analysis for product deployment. Internal audits aim to evaluate how well the product candidate, once in real-world operation, will fit the expected system behaviour encoded in standards. 

A modification in objective from a post-deployment audit to pre-deployment audit applied throughout the development process enables proactive ethical intervention methods, rather than simply informing reactive measures only implementable after deployment, as is the case with a purely external approach. Because there is an increased level of system access in an internal audit, identified gaps in performance or processes can be mapped to sociotechnical considerations that should be addressed through joint efforts with product teams. As the audit results can lead to ambiguous conclusions, it is critical to identify key stakeholders and decision makers who can drive appropriate responses to audit outcomes.

Additionally, with an internal audit, because auditors are employees of the organization and communicate their findings primarily to an internal audience, there is opportunity to leverage these audit outcomes for recommendations of structural organizational changes needed to make the entire engineering development process auditable and aligned with ethical standards. %There is thus opportunity for internal audits to contribute to proactive ethical intervention methods, rather than simply informing reactive measures only implementable after deployment, as is the case with a purely external approach.
Ultimately, internal audits complement external accountability, generating artifacts or transparent information \cite{shah2018algorithmic} that third parties can use for external auditing, or even end-user communication. 
%There is also the need to identify which results and biases are \say{unfixable} social realities, and explore mechanisms in user communication or the deployment environment to work with these immutable elements. Similar to the concept of structural vulnerabilities in medicine, it becomes important to operationalize the assessment of these societal disparities, as well as effectively communicate the expected performance differences in deployment \cite{quesada2011structural}. 
Internal audits can thus enable review and scrutiny from additional stakeholders, by enforcing transparency through stricter reporting requirements.

%\subsubsection{Transparency and accountability.} 

%Similar to work in algorithmic accountability \cite{shah2018algorithmic, amini2019uncovering, whittaker2018ai}, transparency around audit findings is a mechanism for accountability. Making results transparent can enable review and scrutiny from additional stakeholders. Internal auditors' direct access to systems can help extend traditional external auditing paradigms to the potential issues feeding into the system, increasing the level of accountability for system creators.

%\subsubsection{The opportunity exists to set up proactive interventions.} 

\section{Lessons From Auditing Practices in Other Industries}
Improving the governance of artificial intelligence development is intended to reduce the risks posed by new technology. While not without faults, safety-critical and regulated industries such as aerospace and medicine have long traditions of auditable processes and design controls that have dramatically improved safety \cite{teixeira2013design, verma2010reliability}.

\subsection{Aerospace}
Globally, there is one commercial airline accident per two million flights \cite{rodrigues2011commercial}. This remarkable safety record is the result of a joint and concerted effort over many years by aircraft and engine manufacturers, airlines, governments, regulatory bodies, and other industry stakeholders \cite{rodrigues2011commercial}. As modern avionic systems have increased in size and complexity (for example, the Boeing 787 software is estimated at 13 million lines of code \cite{judas2011historical}), the standard 1-in-1,000,000,000 per use hour maximum failure probability for critical aerospace systems remains an underappreciated engineering marvel \cite{driscoll2003byzantine}.

However, as the recent Boeing 737 MAX accidents indicate, safety is never finished, and the qualitative impact of failures cannot be ignored---even one accident can impact the lives of many and is rightfully acknowledged as a catastrophic tragedy. Complex systems tend to drift toward unsafe conditions unless constant vigilance is maintained \cite{leveson2011engineering}. It is the sum of the tiny probabilities of individual events that matters in complex systems---if this grows without bound, the probability of catastrophe goes to one. The \textit{Borel-Cantelli} Lemmas are formalizations of this statistical phenomenon \cite{chung1952application}, which means that we can never be satisfied with safety standards. Additionally, standards can be compromised if competing business interests take precedence. Because the non-zero risk of failure grows over time, without continuous active measures being developed to mitigate risk, disaster becomes inevitable  \cite{haigh2012probability}.   

%It is important to ensure that the assumptions that underpin the original software and hardware system designs are clearly \textit{documented} \cite{hall2014distributed}. Likewise, we argue that  it is important to ensure that the assumptions that underpin the original datasets and models are clearly documented, so that downstream developers know the limitations of datasets and models. This approach is not perfect, but the high stakes in aerospace give rise to a critical auditing infrastructure that can be adopted in AI auditing practices. We detail some of the key components below.

\subsubsection{Design checklists}
Checklists are simple tools for assisting designers in having a  more informed view of important questions, edge cases and failures \cite{hall2014distributed}. Checklists are widely used  in aerospace for their proven ability to improve safety and  designs. There are several cautions about using checklists during the development of complex software, such as the risk of blind application, the broader context and nuanced interrelated concerns are not considered. However, a checklist can be beneficial. It is good practice to avoid yes/no questions to reduce the risk that the checklist becomes a box-ticking activity, for example by asking designers and engineers to describe their processes for assessing ethical risk. Checklist use should also be related to real-world failures and higher-level system hazards.

\subsubsection{Traceability} 
Another key concept from aerospace and safety-critical software engineering is \textit{traceability}---which is concerned with the relationships between product requirements, their sources and system design. This practice is familiar to the software industry in requirements engineering \cite{bennaceur2019requirements}. However, in AI research, it can often be difficult to trace the provenance of large datasets or to interpret the meaning of model weights---to say nothing of the challenge of understanding how these might relate to system requirements. Additionally, as the complexity of sociotechnical systems is rapidly increasing, and as the speed and complexity of large-scale artificial intelligence systems increase, new approaches are necessary to understand risk \cite{leveson2011engineering}.

\subsubsection{Failure Modes and Effects Analysis}\label{sec:fmea}
Finally, a standard tool in safety engineering is a \textit{Failure Modes and Effects Analysis} (FMEA), methodical and systematic risk management approach that examines a proposed design or technology for foreseeable failures  \cite{stamatis2003failure}. The main purpose of a FMEA is to define, identify and eliminate potential failures or problems in different products, designs, systems and services. Prior to conducting a FMEA, known issues with a proposed technology should be thoroughly mapped through a literature review and by collecting and documenting the experiences of the product designers, engineers and managers. Further, the risk exercise is based on known issues with relevant datasets and models, information that can be gathered from interviews and from extant technical documentation.

FMEAs can help designers improve or upgrade their products to reduce risk of failure. They can also help decision makers formulate corresponding preventive measures or improve reactive strategies in the event of post-launch failure. FMEAs are widely used in many fields including aerospace, chemical engineering, design, mechanical engineering and medical devices. To our knowledge, however, the FMEA method has not been applied to examine ethical risks in production-scale artificial intelligence models or products.

\subsection{Medical devices} 

Internal and external quality assurance audits are a daily occurrence in the pharmaceutical and medical device industry. Audit document trails are as important as the drug products and devices themselves. The history of quality assurance audits in medical devices dates from several medical disasters in which devices, such as infusion pumps and autoinjectors, failed or were used improperly \cite{vanderveen2005averting}.   

\subsubsection{Design Controls} For medical devices, the stages of product development are strictly defined. In fact, federal law (Code of Federal Regulations Title 21) mandates that medical-device makers establish and maintain \say{design control} procedures to ensure that design requirements are met and designs and development processes are auditable. Practically speaking, design controls are a documented method of ensuring that the end product matches the intended use, and that potential risks from using the technology have been anticipated and mitigated \cite{teixeira2013design}. The purpose is to ensure that anticipated risks related to the use of technology are driven down to the lowest degree that is reasonably practicable.

\subsubsection{Intended Use}
Medical-device makers must maintain procedures to ensure that design requirements meet the \say{intended use} of the device. The intended use of a \say{device} (or, increasingly in medicine, an algorithm---see \cite{price2017regulating} for more) determines the level of design control required: for example, a tongue depressor (a simple piece of wood) is the lowest class of risk (Class I), while a deep brain implant would be the highest (Class III). The intended use of a tongue depressor could be \say{to displace the tongue to facilitate examination of the surrounding organs and tissues}, differentiating a tongue depressor from a Popsicle stick. This may be important when considering an algorithm that can be used to identify cats or to identify tumors; depending on its intended use, the same algorithm might have drastically different risk profiles, and additional risks arise from unintended uses of the technology.

\subsubsection{Design History File}\label{sec:designhistoryfile}
For products classified as medical devices, at every stage of the development process, device makers must document the design input, output, review, verification, validation, transfer and changes---the design control process (section 3.2.1). Evidence that medical device designers and manufacturers have followed design controls must be kept in a design history file (DHF), which must be an accurate representation and documentation of the product and its development process. Included in the DHF is an extensive risk assessment and hazard analysis, which must be continuously updated as new risks are discovered. Companies also proactively maintain \say{post-market surveillance} for any issues that may arise with safety of a medical device. 

\subsubsection{Structural Vulnerability}
In medicine there is a deep acknowledgement of socially determinant factors in healthcare access and effectiveness, and an awareness of the social biases influencing the dynamic of prescriptions and treatments. This widespread acknowledgement led to the framework of operationalizing structural vulnerability in healthcare contexts, and effectively the design of an assessment tool to record the anticipated social conditions surrounding a particular remedy or medical recommendation \cite{quesada2011structural}. Artificial intelligence models are equally subject to social influence and social impact, and undergoing such assessments on more holistic and population- or environment-based considerations is relevant to algorithmic auditing.

\subsection{Finance}
As automated accounting systems started to appear in the 1950s, corporate auditors continued to rely on manual procedures to audit \say{around the computer}. In the 1970s, the Equity Funding Corporation scandal and the passage of the Foreign Corrupt Practices Act spurred companies to more thoroughly integrate internal controls throughout their accounting systems. This heightened the need to audit these systems directly. The 2002 Sarbanes-Oxley Act introduced sweeping changes to the profession in demanding greater focus on financial reporting and fraud detection \cite{byrnes2018evolution}.

Financial auditing had to play catch-up as the complexity and automation of financial business practices became too unwieldy to manage manually. Stakeholders in large companies and government regulators desired a way to hold companies accountable. Concerns among regulators and shareholders that the managers in large financial firms would squander profits from newly created financial instruments prompted the development of financial audits  \cite{styhre2015financialization}.

Additionally, as financial transactions and markets became more automated, abstract and opaque, threats to social and economic values were answered increasingly with audits. But financial auditing lagged behind the process of technology-enabled financialization of markets and firms. 

\subsubsection{Audit Infrastructure}
In general, internal financial audits seek assurance that the organization has a formal governance process that is operating as intended: values and goals are established and communicated, the accomplishment of goals is monitored, accountability is ensured and values are preserved. Further, internal audits seek to find out whether significant risks within the organization are being managed and controlled to an acceptable level \cite{soh2011internal}.

Internal financial auditors typically have unfettered access to necessary information, people, records and outsourced operations across the organization. IIA Performance Standard 2300, Performing the Engagement \cite{institute2007professional}, states that internal auditors should identify, analyze, evaluate and record sufficient information to achieve the audit objectives. The head of internal audit determines how internal auditors carry out their work and the level of evidence required to support their conclusions.

\subsection{Discussion and Challenges}
The lessons from other industries above are a useful guide toward building internal accountability to society as a stakeholder. Yet, there are many novel and unique aspects of artificial intelligence development that present urgent research challenges to overcome. 

Current software development practice in general, and artificial intelligence development in particular, does not typically follow the \textit{waterfall} or verification-and-validation approach \cite{cusumano1995beyond}. These approaches are still used, in combination with agile methods, in the above-mentioned industries because they are much more documentation-oriented, auditable and requirements-driven. Agile artificial intelligence development is much faster and iterative, and thus presents a challenge to auditability. However, applying agile methodologies to internal audits themselves is a current topic of research in the internal audit profession.\footnote{https://deloitte.wsj.com/riskandcompliance/2018/08/06/mind-over-matter-implementing-agile-internal-audit/}

Most internal audit functions outside of heavily regulated industries tend to take a risk-based approach. They work with product teams to ask "what could go wrong" at each step of a process and use that to build a risk register \cite{patterson2002risk}. This allows risks to rise to the surface in a way that is informed by the people who know these processes and systems the best. Internal audits can also leverage relevant experts from within the company to facilitate such discussions and provide additional insight on potential risks \cite{bing2005allocation}.

Large-scale production AI systems are extraordinarily complex, and a critical line of future research relates to addressing the interaction of highly complex coupled sociotechnical systems. Moreover, there is a dynamic complex interaction between users as sources of data, data collection, and model training and updating. Additionally, governance processes based solely on risk have been criticized for being unable to anticipate the most profound impacts from technological innovation, such as the financial crisis in 2008, in which big data and algorithms played a large role  \cite{muniesa2013responsible, noble2018algorithms, o2016weapons}.

With artificial intelligence systems it can be difficult to trace model output back to requirements because these may not be explicitly documented, and issues may only become apparent once systems are released. However, from an ethical and moral perspective it is incumbent on producers of artificial intelligence systems to anticipate ethics-related failures before launch. However, as \cite{parker2012unexpected} and \cite{holstein2018improving} point out, the design, prototyping and maintenance of AI systems raises many unique challenges not commonly faced with other kinds of intelligent systems or computing systems more broadly. For example, \textit{data entanglement} results from the fact that artificial intelligence is a tool that mixes data sources together. As Scully et al.~point out, artificial intelligence models create entanglement and make the isolation of improvements effectively impossible \cite{sculley2014machine}, which they call \textit{Change Anything Change Everything}. We suggest that by having explicit documentation about the purpose, data, and model space, potential hazards could be identified earlier in the development process. 

Selbst and Barocas argue that \say{one must seek explanations of the process behind a model`s development, not just explanations of the model itself} \cite{selbst2018intuitive}. As a relatively young community focused on fairness, accountability, and transparency in AI, we have some indication of the system culture requirements needed to normalize, for example, an adequately thorough documentation procedure and guidelines \cite{gebru2018datasheets, mitchell2019model}. Still, we lack the formalization of a standard model development template or practice, or process guidelines for when and in which contexts it is appropriate to implement certain recommendations. In these cases, internal auditors can work with engineering teams to construct the missing documentation to assess practices against the scope of the audit. Improving documentation can then be a remediation for future work.  

Also, as AI is at times considered a \say{general purpose technology} with multiple and dual uses \cite{trajtenberg2018ai}, the lack of reliable standardization poses significant challenges to governance efforts. This challenge is compounded by increasing customization and variability of what an AI product development lifecycle looks like depending on the anticipated context of deployment or industry.

We thus combine learnings from prior practice in adjacent industries while recognizing the uniqueness of the commercial AI industry to identify key opportunities for internal auditing in our specific context. We do so in a way that is appropriate to the requirements of an AI system. 

\begin{figure*}
  \centering
    \includegraphics[width=\textwidth]{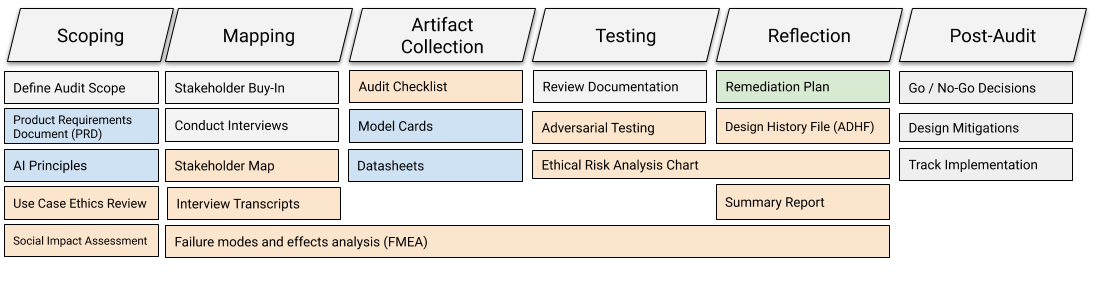}
  \caption{Overview of Internal Audit Framework. Gray indicates a process, and the colored sections represent documents. Documents in orange are produced by the auditors, blue documents are produced by the engineering and product teams and green outputs are jointly developed.}\label{fig:audit2}
  %During the scoping phase, questions such as who will be using the system and for what purpose are addressed. System risks are then identified and explored further as the system is mapped, and documentation on various system elements are collected. The review of this documentation in addition to adversarial testing further informs potential failures of the system and allows a judgment on which risks are more likely to occur. Prioritized risks are then added to a remediation plan, which informs post-audit decision making. An audit report is compiled from document artifacts generated throughout the audit procedure from planning to evaluation review.
%  }
\end{figure*}

\section{SMACTR: An internal audit framework}

We now outline the components of an initial internal audit framework, which can be framed as encompassing five distinct stages---Scoping, Mapping, Artifact Collection, Testing and Reflection (SMACTR)---all of which have their own set of documentation requirements and account for a different level of the analysis of a system. Figure \ref{fig:audit2} illustrates the full set of artifacts recommended for each stage. %Templates and example implementations of the Social Impact Assessment, Audit Checklist, Failure Modes and Effects Analysis (FMEA) and Ethical Risk Analysis are included in the Supplementary Materials for this paper. 
%We identify a minimal set of essential documents required at each stage of the audit, including informational documentation and processes on the technological system \cite{hind2018increasing} and any ethical risk documents that review ethical issues, analyze social impact, or analyze the decisions and biases of the stakeholders contributing to the project. 

To illustrate the utility of this framework, we contextualize our descriptions with the hypothetical example of Company X Inc., a large multinational software engineering consulting firm, specializing in developing custom AI solutions for a diverse range of clients. We imagine this company has designated five AI principles, paraphrased from the most commonly identified AI principles in a current online English survey \cite{jobin2019artificial}--"Transparency", "Justice, Fariness \& Non-Discrimination", "Safety \& Non-Maleficence", "Responsibility \& Accountability" and "Privacy". We also assume that the corporate structure of Company X is typical of any technical consultancy, and design our stakeholder map by this assumption.

Company X has decided to pilot the SMACTR internal audit framework to fulfill a corporate mandate towards responsible innovation practice, accommodate external accountability and operationalize internal consistency with respect to its identified AI principles. The fictional company thus pilots the audit framework on two hypothetical client projects.

The first (hypothetical) client wishes to develop a child abuse screening tool similar to that of the real cases extensively studied and reported on \cite{chouldechova2018case,cuccaro2017risk,goldhaber2019impact,keddell2019algorithmic,courtland2018bias,eubanks2018child}. This complex case intersects heavily with applications in high-risk scenarios with dire consequences. This scenario demonstrates how, for algorithms interfacing with high-risk contexts, a structured framework can allow for the careful consideration of all the possibilities and risks with taking on the project, and the extent of its understood social impact. %This case would also illustrate how an organization could prioritize audit resources and which use cases may be most qualified to necessitate a full audit.

The second invented client is Happy-Go-Lucky, Inc., an imagined photo service company looking for a smile detection algorithm to automatically trigger the cameras in their installed physical photo booths. In this scenario, the worst case is a lack of customer satisfaction---the stakes are low and the situation seems relatively straightforward. This scenario demonstrates how in even seemingly simple and benign cases, ethical consideration of system deployment can reveal underlying issues to be addressed prior to deployment, especially when we contextualize the model within the setting of the product and deployment environment.

An end-to-end worked example of the audit framework is available as supplementary material to this paper for the Happy-Go-Lucky, Inc. client case. This includes demonstrative templates of all recommended documentation, with the exception of specific process files such as any experimental results, interview transcripts, a design history file and the summary report. Workable templates can also be accessed as an online resource \underline{\href{https://drive.google.com/drive/folders/1NUSTvWObrDSkjmWTrDCxKoPkXmypguiv?usp=sharing}{here}}.

\subsection{The Governance Process}
To design our audit procedure, we suggest complementing formal risk assessment methodologies with ideas from responsible innovation, which stresses four key dimensions: \textit{anticipation}, \textit{reflexivity}, \textit{inclusion} and \textit{responsiveness} \cite{stilgoe2013developing}, as well as system-theoretic concepts that help grapple with increasing complexity and coupling of artificial intelligence systems with the external world \cite{leveson2011engineering}. Risk-based assessments can be limited in their ability to capture social and ethical stakes, and they should be complemented by anticipatory questions such as, \say{what if...?}. The aim is to increase ethical foresight through systematic thinking about the larger sociotechnical system in which a product will be deployed \cite{mittelstadt2016ethics}. There are also intersections between this framework and just effective product development theory \cite{brown1995product}, as many of the components of audit design refocus the product development process to prioritize the user and their ultimate well-being, resulting in a more effective product performance outcome.

At a minimum, the internal audit process should enable critical reflections on the potential impact of a system, serving as internal education and training on ethical awareness in addition to leaving what we refer to as a \say{transparency trail} of documentation at each step of the development cycle (see Figure~\ref{fig:audit2}). To shift the process into an actionable mechanism for accountability, we present a validated and transparently outlined procedure that auditors can commit to. The thoroughness of our described process will hopefully engage the trust of audit targets to act on and acknowledge  post-audit recommendations for engineering practices in alignment with prescribed AI principles. 

This process primarily addresses how to conduct internal audits, providing guidance for those that have already deemed an audit necessary but would like to further define the scope and execution details. Though not covered here, an equally important process is determining what systems to audit and why. Each industry has a way to judge what requires a full audit, but that process is discretionary and dependent on a range of contextual factors pertinent to the industry, the organization, audit team resourcing, and the case at hand. Risk prioritization and the necessary variance in scrutiny is a separately interesting and rich research topic on its own. The process outlined below can be applied in full or in a lighter-weight formulation, depending on the level of assessment desired.

\subsection{The Scoping Stage}

%4.2 The Scoping Stage

For both clients, a product or request document is provided to specify the requirements and expectations of the product or feature. The goal of the scoping stage is to clarify the objective of the audit by reviewing the motivations and intended impact of the investigated system, and confirming the principles and values meant to guide product development. This is the stage in which the risk analysis begins by mapping out intended use cases and identifying analogous deployments either within the organization or from competitors or adjacent industries. The goal is to anticipate areas to investigate as potential sources of harm and social impact. At this stage, interaction with the system should be minimal. 

In the case of the smile-triggered phone booth, a smile detection model is required, providing a simple product, with not a broad scope of considerations as the potential for harm does not go much beyond inconvenience or customer exclusion and dissatisfaction. For the child abuse detection product, there are many more approaches to solving the issue and many more options for how the model interacts with the broader system. The use case itself involves many ethical considerations, as an ineffective model may result in serious consequences like death or family separation.

The key artifacts developed by the auditors from this stage include an ethical review of the system use case and a social impact assessment. Pre-requisite documents from the product and engineering team should be a declaration or confirmation statement of ethical objectives, standards and AI principles. The product team should also provide a Product Requirements Document (PRD), or project proposal from the initial planning of the audited product. 

\subsubsection {Artifact: Ethical Review of System Use Case}
When a potential AI system is in the development pipeline, it should be reviewed with a series of questions that first and foremost check to see, at a high level, whether the technology aligns with a set of ethical values or principles. This can take the form of an ethical review that considers the technology from a responsible innovation perspective by asking who is likely to be impacted and how. 

Importantly, we stress standpoint diversity in this process. \textbf{Algorithm development implicitly encodes developer assumptions that they may not be aware of, including ethical and political values.} Thus it is not always possible for individual technology workers to identify or assess their own biases or faulty assumptions \cite{intemann201025}. For this reason, a critical range of viewpoints is included in the review process. The essential inclusion of independent domain experts and marginalized groups in the ethical review process "has the potential to lead to more rigorous critical reflection because their experiences will often be precisely those that are most needed in identifying problematic background assumptions and revealing limitations with research questions, models, or methodologies" \cite{intemann201025}. Another method to elicit implicit biases or motivated cognition \cite{kruglanski1996motivated} is to ask people to reflect on their preliminary assessment and then ask whether they might have reason to regret the action later on. This can shed light on how our position in society biases our assumptions and ways of knowing \cite{dobbe2018broader}.

%It is also important to collect evidence of potential impact from those who are likely to be impacted externally, for example, by algorithmic decisions in hiring. Ethical due diligence and responsible innovation require gathering evidence and understanding potential consequences of the system. This forms the motivation for social impact and human rights assessments outlined below. 
%: whether the project is approved or not, what modifications (if any) are required prior to approval, or if the project crosses any ethical red lines.

An internal ethics review board that includes a diversity of voices should review proposed projects and document its views. Internal ethics review boards are common in biomedical research, and the purpose of these boards is to ensure that the rights, safety, and well-being of all human subjects involved in medical research are protected \cite{general2014world}. Similarly, the purpose of an ethics review board for AI systems includes safeguarding human rights, safety, and well-being of those potentially impacted. 
%These early-stage assessments will provide development guidance based on the risk profile of the system, and at this stage the FMEA (Section \ref{sec:fmea}) should begin and risks prioritized for later testing. As in the medical device space, that entails, for instance, the design controls that should be implemented. 

\subsubsection{Artifact: Social Impact Assessment}
A social impact assessment should inform the ethical review. Social impact assessments are commonly defined as a method to analyze and mitigate the unintended social consequences, both positive and negative, that occur when a new development, program, or policy engages with human populations and communities \cite{vanclay2003international}. In it, we describe how the use of an artificial intelligence system might change people's ways of life, their culture, their community, their political systems, their environment, their health and well-being, their personal and property rights, and their experiences (positive or negative) \cite{vanclay2003international}.
 %Social impact assessments are also understood as a way to assess social change. 

The social impact assessment includes two primary steps: an assessment of the severity of the risks, and an identification of the relevant social, economic, and cultural impacts and harms that an artificial intelligence system applied in context may create. The severity of risk is the degree to which the specific context of the use case is assessed to determine the degree in which potential harms may be amplified. The severity assessment proceeds from the analysis of impacts and harms to give a sense of the relative severity of the harms and impacts depending on the sensitivity, constraints, and context of the use case. 

\subsection{The Mapping Stage}

%4.3 The Mapping Stage

The mapping stage is not a step in which testing is actively done, but rather a review of what is already in place and the perspectives involved in the audited system. This is also the time to map internal stakeholders, identify key collaborators for the execution of the audit, and orchestrate the appropriate stakeholder buy-in required for execution. At this stage, the FMEA (Section \ref{sec:fmea}) should begin and risks should be prioritized for later testing. %A preferred pre-requisite for this stage is that AI principles, which represent high-level ethical ideals and values for the system, should ideally be translated into internal policies and standards. For instance, the principle of transparency can translate to engineering documentation requirements. High-level principles can also be used to help identify risks and provide the room for lower-level definitions of what can be considered a failure to achieve the ideal. The audit team and involved stakeholders can collectively confirm at least a contextual definition of the declared principles. 

%As in the medical device space that entails, for instance, the design controls that should be implemented.

As Company X is a consultancy, this stage mainly requires identifying the stakeholders across product and engineering teams anchored to this particular client project, and recording the nature of their involvement and contribution. This enables an internal record of individual accountability with respect to participation towards the final outcome, and enables the trace of relevant contacts for future inquiry.  

For the child abuse detection algorithm, the initial identification of failure modes reveals the high stakes of the application, and immediate threats to the "Safety \& Non-Maleficence" principle. False positives overwhelm staff and may lead to the separation of families that could have recovered. False negatives may result in a dead or injured child that could have been rescued. %These consequences are irrevocable serious outcomes and immediate threats to the "Safety \& Non-Maleficence" principle. 
For the smile detector, failures disproportionately impact those with alternative emotional expressions---those with autism, different cultural norms on the formality of smiling, or different expectations for the photograph who are then excluded from the product by design.%, jeopardizing the "Justice, Fariness \& Non-Discrimination" principle.

The key artifacts from this stage include a stakeholder map and collaborator contact list, a system map of the product development lifecycle, and the engineering system overview, especially in cases where multiple models inform the end product. Additionally, this stage includes a design history file review of all existing documentation of the development process or historical artifacts on past versions of the product. Finally, it includes a report or interview transcripts on key findings from internal ethnographic fieldwork involving the stakeholders and engineers. 

\subsubsection{Artifact: Stakeholder Map}

Who was involved in the system audit and collaborators in the execution of the audit should be outlined. Clarifying participant dynamics ensures a more transparent representation of the provided information, giving further context to the intended interpretation of the final audit report. 
%This is to maintain transparency about which teams were consulted and which biases or perspectives they may have (including the auditors themselves). A stakeholder map of the involved parties at the organization level illustrating the teams and functions involved in the project helps to trace the decision-making process about specific components or stages of the product development lifecycle. The map can simply be a list of important stakeholders and project engineers that describes their role in system development. 

%Next, it is important to gather buy-in from the many stakeholders involved in the product development process. This involves mapping out a system architecture and the relevant teams involved in building the system. 

\subsubsection{Artifact: Ethnographic Field Study}
%We overview the development process and organizational structure of an internal audit. 

As Leveson points out, bottom-up decentralized decision making can lead to failures in complex sociotechnical systems \cite{leveson2011engineering}. Each local decision may be correct in the limited context in which it was made, but can lead to problems when these decisions and organizational behaviors interact. With modern large-scale artificial intelligence projects and API development, it can be difficult to gain a shared understanding at the right level of system description to understand how local decisions, such as the choice of dataset or model architecture, will impact final system behavior.

Therefore, ethnography-inspired fieldwork methodology based on how audits are conducted in other industries, such as finance \cite{styhre2015financialization} and healthcare \cite{rodriguez2014audits} is useful to get a deeper and qualitative understanding of the engineering and product development process. As in internal financial auditing, access to key people in the organization is important. This access involves semi-structured interviews with a range of individuals close to the development process and documentation gathering to gain an understanding of possible gaps that need to be examined more closely.

Traditional metrics for artificial intelligence like loss may conceal fairness concerns, social impact risks or abstraction errors \cite{selbst2019fairness}. A key challenge is to assess how the numerical metrics specified in the design of an artificial intelligence system reflect or conform with these values. Metrics and measurement are important parts of the auditing process, but should not become aims and ends in themselves when weighing whether an algorithmic system under audit is ethically acceptable for release. Taking metrics measured in isolation risks recapitulating the abstraction error that \cite{selbst2019fairness} point out, "To treat fairness and justice as terms that have meaningful application to technology separate from a social context is therefore to make a category error, or as we posit here, an abstraction error." What we consider data is already an interpretation, highly subjective and contested \cite{furner2016data}. Metrics must be understood in relation to the engineering context in which they were developed and the social context into which they will be deployed. During the interviews, auditors should capture and pay attention to what falls outside the measurements and metrics, and to render explicit the assumptions and values the metrics apprehend \cite{styhre2018unfinished}. For example, the decision about whether to prioritize the false positive rate over false negative rate (precision/recall) is a question about values and cannot be answered without stating the values of the organization, team or even engineer within the given development context. %What is the intended use of the system and who are the intended users? What are the intended use environments?  %The findings of the audit and the audit process itself should \say{guide, prompt and open up space for essential governance discussions aimed at supporting, but not dictating, decisions about the framing, direction, pace and trajectory of contentious and innovative research} \cite{stilgoe2013developing},.

%These early-stage assessments will provide development guidance based on the risk profile of the system, and 

\subsection{The Artifact Collection Stage}

%4.4 The Artifact Collection Stage

Note that the collection of these artifacts advances adherence to the declared AI principles of the organization on "Responsibility \& Accountability" and "Transparency".

%ADD CITES
In this stage, we identify and collect all the required documentation from the product development process, in order to prioritize opportunities for testing. Often this implies a record of data and model dynamics though application-based systems can include other product development artifacts such as design documents and reviews, in addition to systems architecture diagrams and other implementation planning documents and retrospectives. 

At times documentation can be distributed across different teams and stakeholders, or is missing altogether. In certain cases, the auditor is in a position to enforce retroactive documentation requirements on the product team, or craft documents themselves. 

The model card for the smile detection model is the template model card from the original paper \cite{mitchell2019model}. A hypothetical datasheet for this system is filled out using studies on the CelebA dataset, with which the smile detector is built \cite{celeba, merler2019diversity}. In the model card, we identify potential for misuse if smiling is confused for positive affect. From the datasheet for the CelebA dataset, we see that although the provided binary gender labels seem balanced for this dataset (58.1\% female, 42\% male), other demographic details are quite skewed (77.8\% aged 0-45, 22.1\% aged over 46 and 14.2\% lighter-skinned, 85.8\% darker-skinned)\cite{merler2019diversity}.

The key artifact from auditors during this stage is the audit checklist, one method of verifying that all documentation pre-requisites are provided in order to commence the audit. Those pre-requisites can include model and data transparency documentation.

\subsubsection{Artifact: Design Checklist}

This checklist is a method of taking inventory of all the expected documentation to have been generated from the product development cycle. It ensures that the full scope of expected product processes and that the corresponding documentation required to be completed before the audit review can begin are finished. This is also a procedural evaluation of the development process for the system, to ensure that appropriate actions were pursued throughout system development ahead of the evaluation of the final system outcome.

\subsubsection{Artifacts: Datasheets and Model Cards}

Two recent standards can be leveraged to create auditable documentation, model cards and datasheets \cite{mitchell2019model,gebru2018datasheets}. Both model cards and datasheets are important tools toward making algorithmic development and the algorithms themselves more auditable, with the aim of anticipating risks and harms with using artificial intelligence systems. Ideally, these artifacts should be developed and/or collected by product stakeholders during the course of system development. 

%The goal is not to comprehensively describe these approaches, but to illustrate how these approaches can be part of a usable audit process to characterize data and models with a view toward anticipating harms or potential social impact. 

%In machine learning it is common to recycle data and models for new purposes. For many industry machine learning practitioners and engineers, the data has already been collected and labelled, the models have already been trained, and their job is to create new applications or APIs from these existing systems. This flexibility is a key advantage for scaling machine learning, however there can be downsides. For example, an image classifier might be trained on a certain type of images from the web, then tasked with making predictions about different types of images from videos. This can lead to biases or failures that are difficult to foresee.

To clarify the intended use cases of artificial intelligence models and minimize their usage in contexts for which they are not well suited, Mitchell et al.~recommend that released models be accompanied by documentation detailing their performance characteristics \cite{mitchell2019model}, called a {\it model card}. This should include information about how the model was built, what assumptions were made during development, and what type of model behavior might be experienced by different cultural, demographic or phenotypic groups. A model card is also extremely useful for internal development purposes to make clear to stakeholders details about trained models that are included in larger software pipelines, which are parts of internal organizational dynamics, which are then parts of larger sociotechnical logics and processes. A robust model card is key to documenting the intended use of the model as well as information about the evaluation data, model scope and risks, and what might be affecting model performance. 

Model cards are intended to complement "Datasheets for Datasets" \cite{gebru2018datasheets}. Datasheets for machine learning datasets are derived by analogy from the electronics hardware industry, where a datasheet for an electronics component describes its operating characteristics, test results, and recommended uses. A critical part of the datasheet covers the data collection process. This set of questions are intended to provide consumers of the dataset with the information they need to make informed decisions about using the dataset: what mechanisms or procedures were used to collect the data? Was any ethical review process conducted? Does the dataset relate to people? 
%As with model cards, a datasheet should answer a series of questions such as: for what purpose was the dataset created? Who created the dataset? Who funded the creation of the dataset? Next, a datasheet would answer questions about the composition of the dataset: what is in the dataset? Documents? Photos with people? Locations? etc. 

This documentation feeds into the auditors' assessment process.

\subsection{The Testing Stage}

%4.5 The Testing Stage

This stage is where the majority of the auditing team's testing activity is done---when the auditors execute a series of tests to gauge the compliance of the system with the prioritized ethical values of the organization. Auditors engage with the system in various ways, and produce a series of artifacts to demonstrate the performance of the analyzed system at the time of the audit. Additionally, auditors review the documentation collected from the previous stage and begin to make assessments of the likelihood of system failures to comply with declared principles.

High variability in approach is likely during this stage, as the tests that need to be executed change dramatically depending on organizational and system context. Testing should be based on a risk prioritization from the FMEA. 

For the smile detector, we might employ counterfactual adversarial examples designed to confuse the model and find problematic failure modes derived from the FMEA. For the child prediction model, we test performance on a selection of diverse user profiles. These profiles can also be treated for variables that correlate with vulnerable groups to test whether the model has learned biased associations with race or SES.

For the ethical risk analysis chart, we look at the principles and realize that there are immediate risks to the "Privacy" principle---with one case involving juvenile data, which is sensitive, and the other involving face data, a biometric. This is also when it becomes clear that in the smiling booth case, there is disproportionate performance for certain underrepresented user subgroups, thus jeopardizing the "Justice, Fariness \& Non-Discrimination" principle.

The main artifacts from this stage of the auditing process are the results of tests such as adversarial probing of the system and an ethical risk analysis chart. 

\subsubsection{Artifact: Adversarial Testing }

Adversarial testing is a common approach to finding vulnerabilities in both pre-release and post-launch technology, for example in privacy and security testing \cite{brubaker2014using}. In general, adversarial testing attempts to simulate what a hostile actor might do to gain access to a system, or to push the limits of the system into edge case or unstable behavior to elicit very-low probability but high-severity failures. %The key question for adversarial testing of algorithmic systems is: "how hard have you tried to break this before launch?" 

In this process, direct non-statistical testing uses tailored inputs to the model to see if they result in undesirable outputs. These inputs can be motivated by an intersectional analysis, for example where an ML system might produce unfair outputs based on demographic and phenotypic groups that might combine in non-additive ways to produce harm, or over time recapitulate harmful stereotypes or reinforce unjust social dynamics (for example, in the form of opportunity denial). This is distinct from adversarially attacking a model with human-imperceptible pixel manipulations to trick the model into misidentifying previously learned outputs \cite{gu2014towards}, but these approaches can be complementary. This approach is more generally defined---encompassing a range of input options to try in an active attempt to fool the system and incite identified failure modes from the FMEA. 

Internal adversarial testing prior to launch can reveal unexpected product failures before they can impact the real world. Additionally, proactive adversarial testing of already-launched products can be a best practice for lifecycle management of released systems. %For example, considering a facial analysis system, an adversarial test could target known weaknesses based on knowledge of the system performance outlined in the model card and FMEA. If performance is skewed for certain demographics, similar faces from intersectional demographics can be shown to the model to elicit failures and point to potential remediation. 
The FMEA should be updated with these results, and the relative changes to risks assessed.  

\subsubsection{Artifact: Ethical Risk Analysis Chart}

The ethical risk analysis chart considers the combination of the likelihood of a failure and the severity of a failure to define the importance of the risk. Highly likely and dangerous risks are considered the most high-priority threats. Each risk is assigned a severity indication of "high", "mid" and "low" depending on their combination of these features. 

Failure likelihood is estimated by considering the occurrence of certain failures during the adversarial testing of the system and the severity of the risk is identified in earlier stages, from informative processes such as the social impact assessment and ethnographic interviews. 

\subsection{The Reflection Stage}

This phase of the audit is the more reflective stage, when the results of the tests at the execution stage are analyzed in juxtaposition with the ethical expectations clarified in the audit scoping. Auditors update and formalize the final risk analysis in the context of test results, outlining specific principles that may be jeopardized by the AI system upon deployment. This phase will reflect on product decisions and design recommendations that could be made following the audit results.

Additionally, key artifacts at this stage may include a mitigation plan or action plan, jointly developed by the audit and engineering teams, that outlines prioritized risks and test failures that the engineering team is in a position to mitigate for future deployments or for a future version of the audited system. %It is also at this stage that we begin the social impact assessments, looking forward at what other environmental and social factors are influenced by the system's deployment. 

%4.6 The Reflection Stage

For the smile detection algorithm, the decision could be to train a new version of the model on more diverse data before considering deployment, and add more samples of underrepresented populations in CelebA to the training data. It could be decided that the use case does not necessarily define affect, but treats smiling as a favourable photo pose. Design choices for other parts of the product outside the model should be considered---for instance, an opt-in functionality with user permissions required on the screen before applying the model-controlled function, and the default being that the model-controlled trigger is disabled. There could also be an included disclaimer on privacy, assuring users of safe practices for face data storage and consent. Once these conditions are met, Company X could be confident to greenlight developing this product for the client.

For the child abuse detection model---this is a more complex decision. Given the ethical considerations involved, the project may be stalled or even cancelled, requiring further inquiry into the ethics of the use case, and the capability of the team to complete the mitigation plan required to deploy an algorithm in such a high risk scenario.

\subsubsection{Artifact: Algorithmic Use-related Risk Analysis and FMEA}

%Algorithmic accidents, such as instances of algorithmic bias or unanticipated feedback loops, are complex processes involving the entire sociotechnical system. 
The risk analysis should be informed by the social impact assessment and known issues with similar models. Following Leveson's work on safety engineering \cite{leveson2011engineering}, we stress that careful attention must be paid to the distinction between the \textit{designers' mental models} of the artificial intelligence system and the \textit{user's mental model}. The designers' mental models are an idealization of the artificial intelligence system before the model is released. Significant differences exist between this ideal model and how the actual system will behave or be used once deployed. This may be due to many factors, such as distributional drift \cite{lehman2019evolutionary} where the training and test set distributions differ from the real-world distribution, or intentional or unintentional misuse of the model for purposes other than those for which it was designed. Reasonable and foreseeable misuse of the model should be anticipated by the designer. Therefore, the \textit{user's mental model} of the system should be anticipated and taken into consideration. Large gaps between the \textit{intended} and \textit{actual} uses of algorithms have been found in contexts such as criminal justice and web journalism \cite{christin2017algorithms}. 

This adds complexity to anticipated hazards and risks, nevertheless these should be documented where possible. Christin points out \say{the importance of studying the practices, uses, and implementations surrounding algorithmic technologies. Intellectually, this involves establishing new exchanges between literatures that may not usually interact, such as critical data studies, the sociology of work, and organizational analysis}. We propose that known use-related issues with deployed systems be taken into account during the design stage. The format of the risk analysis can be variable depending on context, and there are many valuable templates to be found in \textit{Failure Modes and Effects Analysis} (Section \ref{sec:fmea}) framing and other risk analysis tools in finance and medical deployments.

\subsubsection{Artifact: Remediation and Risk Mitigation Plan} 
After the audit is completed and findings are presented to the leadership and product teams, it is important to develop a plan for remediating these problems. The goal is to drive down the risk of ethical concerns or potential negative social impacts to the extent reasonably practicable. This plan can be reviewed by the audit team and leadership to better inform deployment decisions. 

For the concerns raised in any audit against ethical values, a technical team will want to know: what is the threshold for acceptable performance? If auditors discover, for example, unequal classifier performance across subgroups, how close to parity is necessary to say the classifier is acceptable? In safety engineering, a risk threshold is usually defined under which the risk is considered tolerable. Though a challenging problem, similar standards could be established and developed in the ethics space as well.

 \subsubsection{Artifact: Algorithmic Design History File}

Inspired by the concept of the design history file from the medical device industry \cite{teixeira2013design}, we propose an algorithmic design history file (ADHF) which would collect all the documentation from the activities outlined above related to the development of the algorithm. It should point to the documents necessary to demonstrate that the product or model was developed in accordance with an organization's ethical values, and that the benefits of the product outweigh any risks identified in the risk analysis process. 

 This design history file would form the basis of the final audit report, which is a written evaluation by the organization's audit team. The ADHF should assist with an audit trail, enabling the reconstruction of key decisions and events during the development of the product. The algorithmic report would then be a distillation and summary of the ADHF.

\subsubsection{Artifact: Algorithmic Audit Summary Report}

The report aggregates all key audit artifacts, technical analyses and documentation, putting this in one accessible location for review. This audit report should be compared qualitatively and quantitatively to the expectations outlined in the given ethical objectives and any corresponding engineering requirements. 

\section{Limitations of Internal Audits}

%In addition to the sociotechnical complexities inherent in artificial intelligence outlined above, one objection might be raised that internal auditors, as employees of the organization being audited, lack the independent and critical perspective to carry out an objective audit. However, we would point to  

Internal auditors necessarily share an organizational interest with the target of the audit. While it is important to maintain an independent and objective viewpoint during the execution of an audit, we awknowledge that this is challenging. The audit is never isolated from the practices and people conducting the audit, just as artificial intelligence systems are not independent of their developers or of the larger sociotechnical system. Audits are not unified or monolithic processes with an objective "view from nowhere", but must be understood as a "patchwork of coupled procedures, tools and calculative processes" \cite{styhre2015financialization}. To avoid audits becoming simply acts of reputation management for an organization, the auditors should be mindful of their own and the organizations' biases and viewpoints. Although long-standing internal auditing practices for quality assurance in the financial, aviation, chemical, food, and pharmaceutical industries have been shown to be an effective means of controlling risk in these industries \cite{taylor2018quality}, the regulatory dynamics in these industries suggest that internal audits are only one important aspect of a broader system of required quality checks and balances.

\section{Conclusion}
%The contribution of this framework is the novel combination of critical social science with safety-engineering approaches into a method for translating AI ethics into practice. 
%The formalization of internal auditing for AI systems through the SMACTR framework %this process establishes the integrity of the internal audit, 
%ensures procedural documentation checkpoints and a transparent presentation of 
%introduces methodology for 

AI has the potential to benefit the whole of society, however there is currently an inequitable risk distribution such that those who already face patterns of structural vulnerability or bias disproportionately bear the costs and harms of many of these systems. Fairness, justice and ethics require that those bearing these risks are given due attention and that organizations that build and deploy artificial intelligence systems internalize and proactively address these social risks as well, being seriously held to account for system compliance to declared ethical principles. 

%As outlined, there are a plethora of developed and powerful tools such as data sheets, model cards, and FactSheets \cite{DBLP:journals/corr/abs-1808-07261} which can be used in more standardized ways by anyone to inspect training data, evaluate models and examine human or social concepts inside complex models. 
%Additionally, future research on interpretability methods such as Testing with Concept Activation Vectors \cite{kim2017interpretability} might be able to provide useful guidance for understanding model failures, or whether a model is using protected attributes in ways that it should not. 

%\section{Acknowledgments}

\bibliographystyle{ACM-Reference-Format}
\bibliography{AuditPaper}

%%
%% If your work has an appendix, this is the place to put it.

\end{document}